\documentclass[acmlarge]{acmart}

\makeatletter
\newcommand{\confshort}{\acmConference@shortname}
\newcommand{\conffull}{\acmConference@name}
\newcommand{\confdate}{\acmConference@date}
\newcommand{\confloc}{\acmConference@venue}
\AtBeginDocument{
  \fancypagestyle{firstpagestyle}{
    \fancyhead{}%
    \fancyfoot[C]{}%
  }
  \fancyhf{}
  \fancyhead[LO]{\@headfootfont\shorttitle}%
  \fancyhead[RE]{\@headfootfont\@shortauthors}%
  \fancyhead[LE]{\@headfootfont\footnotesize \confshort, \confdate, \confloc}%
  \fancyhead[RO]{\@headfootfont\footnotesize \confshort, \confdate, \confloc}%
  \fancyfoot[C]{}%
}
\makeatother

\copyrightyear{2026}
\acmYear{2026}
\setcopyright{cc}
\setcctype{by}
\acmConference[FAccT '26]{The 2026 ACM Conference on Fairness, Accountability, and Transparency}{June 25--28, 2026}{Montreal, QC, Canada}
\acmBooktitle{The 2026 ACM Conference on Fairness, Accountability, and Transparency (FAccT '26), June 25--28, 2026, Montreal, QC, Canada}
\acmDOI{}
\acmISBN{}

\usepackage{listings}

\usepackage{booktabs}
\usepackage{tabularx}
\usepackage{array}
\usepackage{ragged2e}
\usepackage{csquotes}
\usepackage{tikz}
\usetikzlibrary{arrows.meta,positioning,shapes.misc,shapes,fit,calc}
\usepackage{adjustbox}
\usepackage{cleveref}
\usepackage{enumitem}

\newcolumntype{P}[1]{>{\raggedright\arraybackslash}p{#1}}
\newcolumntype{Y}{>{\raggedright\arraybackslash}X}

\newcommand{\RightToAI}{Right to AI}
\newcommand{\DelibCell}{deliberative cell}
\newcommand{\DelibCells}{deliberative cells}
\newcommand{\ValueRegister}{value register}

\newcommand{\ProjA}{Street Review}
\newcommand{\ProjB}{LIVS}
\title[Pluralistic-Alignment Urbanism]{Pluralistic-Alignment Urbanism: Operationalizing a Right to AI for Inclusive Public Space}

\author{Rashid Mushkani}
\affiliation{
  \institution{Université de Montréal},
  \institution{Mila -- Québec AI Institute}
  \city{Montr\'eal}
  \country{Canada}
}
\email{rashidmushkani@gmail.com}
\renewcommand{\shortauthors}{Mushkani}

\begin{document}

\begin{abstract}
Municipal agencies increasingly use machine learning to inventory sidewalks, score streetscapes, and generate visualizations of public-space interventions.
These systems produce scores, maps, and synthetic imagery that enter budgeting, design iteration, and public justification.
Because judgments about inclusion, safety, and belonging remain contested in public space, municipal AI governance cannot rely on a single evaluation target as if it were neutral or universally shared. This paper proposes Pluralistic-Alignment Urbanism (PAU), a procedural governance framework that treats public-space AI systems as civic infrastructure and formulates a procedural \RightToAI{} for municipal uses of such systems.
We use two participatory case studies to examine what disagreement, subgroup variation, bounded predictive scaling, and neutral preference judgments can realistically support in municipal practice.
PAU organizes governance around entitlements, municipal duties, documentation artifacts, and recourse triggers for systems that represent public space or generate planning imagery.
The cases are grounded in participatory collaborations with community organizations in Montr\'eal, Canada.
\ProjA{} elicits resident criteria for streetscape evaluation and trains a subgroup-aware scaling model that maps co-produced judgments; the model attains $R^2=0.89$ on a held-out test set.
Here, train/test evaluation is used to assess the bounded feasibility of scaling co-produced judgments into auditable planning artifacts, rather than to assert a single correct target for inclusive streets.
\ProjB{} (a \textit{Local Intersectional Visual Spaces} dataset) constructs pluralistic preference data for aligning text-to-image models and treats neutral selections as evidence of indeterminacy. In an evaluation collected after preference tuning (2{,}100 comparisons), neutral selections are 52.4\% (bootstrap 95\% CI [50.2, 54.5]), and Direct Preference Optimization (DPO)-tuned outputs are preferred in 33.3\% ([31.3, 35.3]). Across the cases, disagreement appears structured, deliberation changes what counts as evidence, scaling is feasible but limited by modality and data coverage, and neutrality in generative evaluation constrains what preference tuning can justify. We translate these constraints into a municipal governance architecture with disaggregated reporting, a versioned \ValueRegister{} that distinguishes benchmarkable from contested dimensions, standing \DelibCells{}, procurement clauses, and defined pause and rollback authority.
\end{abstract}



\maketitle

\paragraph{Contributions.}
This paper proposes Pluralistic-Alignment Urbanism (PAU), a procedural governance framework for public-space AI that treats persistent disagreement as a condition of legitimate municipal decision-making rather than as noise to be eliminated \citep{Mouffe2000,Galston2002,AroyoWelty2015,PavlickKwiatkowski2019}. It synthesizes two participatory case studies to show how disagreement, deliberation, bounded predictive scaling, and indeterminacy constrain what such systems can legitimately represent and justify in practice \citep{EspelandStevens2008,SelbstEtAl2019}. It then translates these constraints into an implementable municipal governance architecture centered on disaggregated reporting, a versioned \ValueRegister{}, standing \DelibCells{}, procurement requirements, and explicit pause and rollback authority \citep{Reisman2018,Selbst2021,Sculley2015,Tabassi2023}.

\section{Introduction}
\label{sec:intro}

Municipal agencies increasingly use machine learning to produce operational representations of streets and public space.
Examples include crowdsourced inventories of sidewalk accessibility barriers \citep{Saha2019}, street-view models that score perceived safety and related streetscape qualities \citep{Naik2014,Kang2023}, and synthetic imagery used to prototype design alternatives or to populate consultation materials \citep{Dubey2024,vonBrackelSchmidt2024GenAIParticipation}.
These systems influence budgeting, prioritization, and public justification by turning heterogeneous experiences into artifacts that appear comparable across space.

In many cities, decisions about public space operate through dual infrastructure: material interventions and an epistemic and algorithmic layer composed of data sources, classification schemes, models, and interfaces \citep{Kitchin2014,Kitchin2014DataRevolution,Batty2018}.
This layer shapes what becomes legible to administrative processes, what can be treated as evidence, and which forms of accountability are available \citep{Star1999,BowkerStar1999,Scott1998,Porter1995}.
Measurement and model outputs can stabilize organizational routines while converting contested values into commensurable quantities \citep{Porter1995,EspelandStevens2008}.
When these representations are treated as objective, the discretion embedded in criteria definitions, aggregation rules, and evaluation protocols becomes difficult to contest.

Public space is governed under persistent disagreement about inclusion, safety, and belonging \citep{Lefebvre1968,Low2000,Mouffe2000}.
Pluralism is an empirical condition because evaluations of the same streetscape vary across social positions, including in intersectional ways \citep{crenshaw1989,low2020social}.
Pluralism is also a governance condition because municipal institutions must justify trade-offs that cannot be resolved by optimizing a single target.

We use the term \emph{evidentiary object} to denote the coupled artifact that enters decision-making, including the stimulus set, criteria definitions, evaluation protocol, aggregation rule, and resulting representation (for example, a score, map, or generated image).
On this view, disagreement, deliberation effects, and neutrality are properties of the evidentiary object and therefore inputs to governance rather than errors to be eliminated \citep{AroyoWelty2015,PavlickKwiatkowski2019}.

This paper defines Pluralistic-Alignment Urbanism (PAU), a governance approach for cities in which public-space decisions depend on both material interventions and algorithmic representations.
PAU operationalizes a procedural \RightToAI{} for public-space AI systems by specifying entitlements, municipal duties, governance artifacts, and recourse mechanisms \citep{Reisman2018,Selbst2021,kaminski2021right,Mushkani2025RightToAI}. We use \ProjA{} and \ProjB{} as empirical case studies through which to ask a narrower question than “does the model perform”: what kinds of municipal claims become legitimate, and what kinds remain out of scope, once disagreement and indeterminacy are treated as decision-relevant evidence? On that basis, we derive a governance architecture with disaggregated reporting, a versioned value register, standing deliberative cells, procurement clauses, and lifecycle checkpoints that define pause and rollback authority \citep{Sculley2015,Tabassi2023}.

\section{Related work}
\label{sec:related}

Three literatures motivate PAU and also reveal the gap this paper addresses: work on AI as urban infrastructure and quantification, work on inclusion and pluralism in public space, and work on participatory and pluralistic AI evaluation. Read together, these literatures show why public-space AI is politically consequential. They do not yet specify how a municipality should treat disagreement and indeterminacy as decision-relevant evidence across commissioning, evaluation, and maintenance.

\subsection{AI as urban infrastructure and the politics of measurement}

Data-driven urbanism expands the use of computational representations in municipal decision-making \citep{Kitchin2014DataRevolution,Townsend2013,SheltonZookWiig2015,batty2024computable}.
Infrastructure scholarship emphasizes that categories, standards, and information systems stabilize organizational routines and distribute power by shaping what becomes visible and actionable \citep{Star1999,BowkerStar1999}.
In public governance, measurement can convert contested values into commensurable indicators and then treat the resulting quantities as objective facts, reducing the space in which value conflict is negotiated \citep{Porter1995,Desrosieres1998,EspelandStevens2008,Merry2016}.
Prior work in algorithmic accountability and critical AI studies has shown how technical systems embed normative assumptions, produce disparate impacts, and evade accountability when design choices are treated as neutral \citep{Pasquale2015,Noble2018,Eubanks2018,SelbstEtAl2019,Crawford2021}. Recent work on perceptive visual urban analytics sharpens this point for municipal settings by showing that visually predictive systems can rely on spurious correlations and remain misaligned with domain knowledge even when benchmark performance looks strong \citep{Alpherts2024Perceptive}. These literatures motivate treating the epistemic and algorithmic layer of urbanism as a governance object rather than as a decision-support tool.

\subsection{Public space and pluralism}

Public space is a political claim about participation, appropriation, and collective life \citep{Lefebvre1968,Harvey2003,Mitchell2003,LowSmith2006}.
Inclusion depends on who feels entitled to be present, who is protected or exposed, and which practices are treated as legitimate \citep{Low2000,VarnaTiesdell2010,Mehta2014}.
Urban design and governance scholarship documents how management practices can produce exclusion through surveillance, privatization, and uneven enforcement \citep{Low2000,Mitchell1995,Brayne2017}.
Pluralism is empirical because evaluations of the same streetscape vary across social positions, including in intersectional ways \citep{crenshaw1989,low2020social}.
Pluralism is also normative because disagreement about what makes public space acceptable is a feature of democratic decision-making \citep{Berlin1969,Galston2002,Mouffe2000}.

\subsection{Participation and power}

Planning and HCI research distinguishes between consultation and decision authority and shows how participation designs can produce procedural legitimacy without transferring power \citep{Arnstein1969,Healey1997,Fung2004,costanzaChock2020}.
STS work on co-production argues that knowledge-making and social order are mutually constituted, so governance requires institutions that attach participation to the production of evidence and standards \citep{Jasanoff2004}.
Recent syntheses of the participatory turn in AI design and studies of public-sector adoption show that stakeholder involvement often arrives after objectives, procurement choices, and evaluation rules are already set, which limits its ability to change consequential design decisions \citep{Delgado2023ParticipatoryTurn,Kawakami2024StudyingUp}.
AI governance scholarship emphasizes rights, contestability, and the limits of voluntary ethics frameworks \citep{Reisman2018,Selbst2021,kaminski2021right,CohenSuzor2024}.
Documentation practices such as datasheets and model cards support auditability by recording intended uses, design choices, and limitations \citep{Gebru2021,Mitchell2019}.

\subsection{Pluralistic annotation and alignment}

Crowdsourcing and evaluation research shows that disagreement can reflect systematic differences in interpretation rather than noise \citep{AroyoWelty2015,PavlickKwiatkowski2019}.
Structured argumentation systems such as Cicero further show that deliberation can improve difficult judgments without implying that a single latent target exists independently of procedure \citep{Chen2019Cicero}.
In alignment research, preference data and reinforcement learning from human feedback can improve model behavior under specified objectives but can also collapse heterogeneity into a single optimized target \citep{Christiano2017,ouyang2022,sorensen2024roadmap}.
Direct Preference Optimization (DPO) provides a computationally simple approach to preference tuning but inherits the governance problem of which preferences are treated as authoritative and how indeterminacy is represented \citep{rafailovDPO2024,wallaceDiffusionDPO2023}.
Adjacent work in urban design and civic engagement examines image-generative AI for co-design, NLP-mediated civic participation, and civic GenAI workflows \citep{Guridi2025PublicParks,Guridi2025ThoughtfulAdoption,Williams2024People,Huang2026UrbanDesign,Shao2025Landscape}. Related critiques also caution that perceptive visual urban analytics can look policy-ready while remaining poorly matched to municipal requirements for traceability, domain fit, and trustworthy use \citep{Alpherts2024Perceptive}.
Taken together, these literatures establish the politics of urban measurement, participation frictions, and plural annotation, but they leave underspecified how municipalities should treat disagreement and indeterminacy as decision-relevant evidence across commissioning, evaluation, and maintenance. PAU fills this gap by linking empirical constraints from representational and generative systems to enforceable procedures, artifacts, and recourse.

\section{Pluralistic-Alignment Urbanism and the procedural Right to AI}
\label{sec:pau-right-to-ai}

PAU integrates three premises.
First, urbanism operates through dual infrastructure: material interventions and an epistemic and algorithmic layer through which conditions are measured and made actionable.
Second, pluralism is a governance condition for public space.
Third, legitimate governance of public-space AI systems requires enforceable procedures and institutions rather than performance claims.

PAU uses \emph{pluralistic alignment} to refer to evaluation and tuning practices that preserve heterogeneous judgments and explicitly represent indeterminacy, rather than collapsing them into a single optimized target.
In preference-based evaluation, PAU treats \emph{neutrality as signal}: a neutral selection is recorded as evidence that a comparison does not support a determinate ordering under a criterion, either because the criterion is context-dependent or because the modality does not provide sufficient evidence.
Two additional working terms recur throughout the paper: the \ValueRegister{} is the versioned municipal record that distinguishes benchmarkable from contested dimensions, and a \DelibCell{} is the standing participatory body that updates that record and can recommend pause, revision, or re-commissioning.

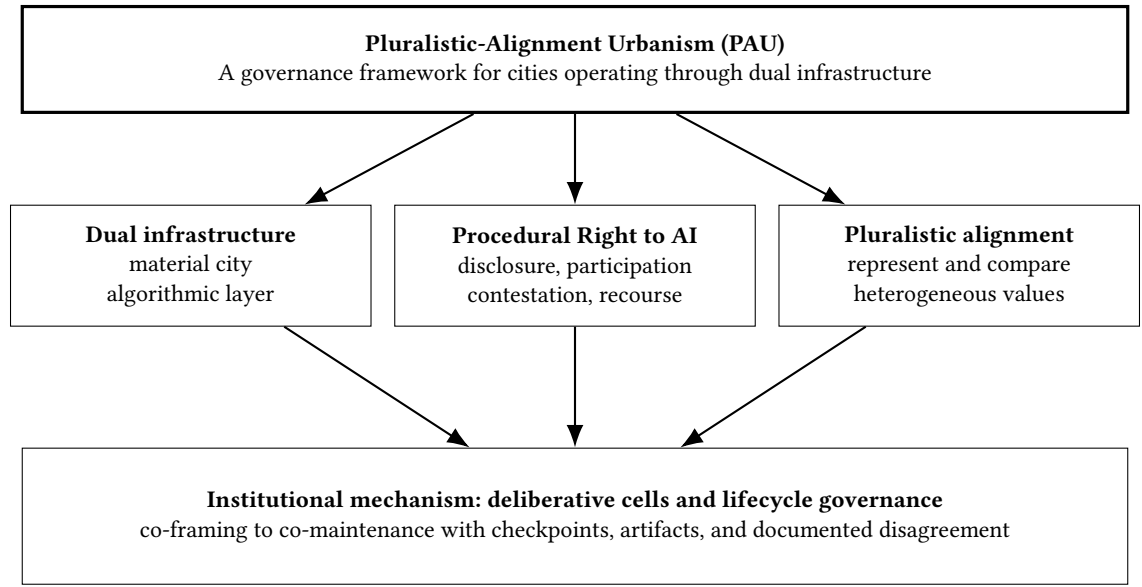
\begin{figure}[t]
\centering
\begin{tikzpicture}[font=\small, node distance=12mm]

\def\PAUwide{0.92\linewidth}
\def\PAUnarrow{0.30\linewidth}

\node[draw, very thick, align=center, minimum width=\PAUwide, minimum height=14mm] (pau) {%
\textbf{Pluralistic-Alignment Urbanism (PAU)}\\
A governance framework for cities operating through dual infrastructure};

\node[draw, align=center, minimum width=\PAUnarrow, minimum height=16mm, below=12mm of pau] (rights) {%
\textbf{Procedural \RightToAI{}}\\
disclosure, participation\\
contestation, recourse};

\node[draw, align=center, minimum width=\PAUnarrow, minimum height=16mm, left=3mm of rights] (dual) {%
\textbf{Dual infrastructure}\\
material city\\
algorithmic layer};

\node[draw, align=center, minimum width=\PAUnarrow, minimum height=16mm, right=3mm of rights] (plural) {%
\textbf{Pluralistic alignment}\\
represent and compare\\
heterogeneous values};

\node[draw, align=center, minimum width=\PAUwide, minimum height=18mm, below=16mm of rights] (inst) {%
\textbf{Institutional mechanism: \DelibCells{} and lifecycle governance}\\
co-framing to co-maintenance with checkpoints, artifacts, and documented disagreement};

\draw[-{Latex[length=3mm]}, thick] (pau) -- (dual);
\draw[-{Latex[length=3mm]}, thick] (pau) -- (rights);
\draw[-{Latex[length=3mm]}, thick] (pau) -- (plural);

\draw[-{Latex[length=3mm]}, thick] (dual) -- (inst);
\draw[-{Latex[length=3mm]}, thick] (rights) -- (inst);
\draw[-{Latex[length=3mm]}, thick] (plural) -- (inst);

\end{tikzpicture}
\caption{PAU treats public-space AI systems as part of an algorithmic layer that requires rights-based governance and pluralistic alignment. Lifecycle governance through \DelibCells{} connects plural evaluations to revisable standards and documented scope limits.}
\Description{Diagram showing PAU as a governance framework connecting dual infrastructure, a procedural Right to AI, pluralistic alignment, and an institutional mechanism of deliberative cells and lifecycle governance.}
\label{fig:pau}
\end{figure}

\subsection{Scope and limits}

In this paper, a procedural \RightToAI{} frames public-space AI systems as governance objects that require democratic authorization and enforceable procedural floors.
The construct draws on rights-based accounts of legitimacy and contestability in algorithmic governance \citep{Reisman2018,Selbst2021,kaminski2021right} and on planning scholarship that treats participation as a design problem that must allocate authority and resources rather than only solicit feedback \citep{Arnstein1969,Healey1997,Fung2004}.
The scope is civic in the sense that it targets systems that shape collective conditions and public justifications, including representational systems that score or map streets and generative systems that produce scenarios for consultation \citep{Mattern2017CityNotComputer,Dubey2024}.

The \RightToAI{} proposed here is not a general right to access AI services, nor a claim that transparency alone or participatory data collection can legitimate municipal AI. It is an administrative construct specifying minimum procedures for public-space AI systems that enter decision processes, procurement, and public justification.

\subsection{Entitlements and recourse}

The \RightToAI{} becomes implementable when entitlements imply municipal duties, duties require governance artifacts, and artifacts enable contestation and revision.
Figure~\ref{fig:right-to-ai-layer} summarizes this translation.

\begin{figure}[t]
\centering
\begin{adjustbox}{width=\linewidth}
\begin{tikzpicture}[
box/.style={draw, rounded corners, align=left, inner sep=10pt, text width=0.28\linewidth},
arrow/.style={-Latex, thick},
font=\small,
node distance=14pt and 16pt
]
\node[box] (rights) {\textbf{\RightToAI{}}\\
Minimum entitlements\\
Bright-line prohibitions\\
Mandated procedures};

\node[box, right=of rights] (duties) {\textbf{Municipal duties}\\
Disclosure and notice\\
Participation as authority\\
Auditability and documentation\\
Recourse and standing\\
Data stewardship};

\node[box, right=of duties] (artifacts) {\textbf{Governance artifacts}\\
Public Space AI Charter\\
\ValueRegister{} (versioned)\\
Model card and data documentation\\
Evaluation logs and audits\\
Disagreement log};

\node[box, right=of artifacts] (recourse) {\textbf{Recourse and re-commissioning}\\
Complaint intake\\
Review and appeal\\
Pause and rollback triggers\\
Revision of standards\\
Decommissioning};

\draw[arrow] (rights) -- (duties);
\draw[arrow] (duties) -- (artifacts);
\draw[arrow] (artifacts) -- (recourse);

\node[
anchor=north west,
text width=\textwidth,
align=center,
font=\footnotesize
] at ([yshift=-14pt]current bounding box.south west) {Entitlements imply duties. Duties require artifacts. Artifacts enable contestation, revision, and accountability over time.};
\end{tikzpicture}
\end{adjustbox}
\caption{The \RightToAI{} as a municipal governance layer connecting entitlements, duties, artifacts, and recourse.}
\Description{Four boxes connected by arrows: Right to AI entitlements lead to municipal duties, which require governance artifacts, which enable recourse and re-commissioning.}
\label{fig:right-to-ai-layer}
\end{figure}

Within this layer, pluralistic representation is a procedural requirement.
Public-space AI systems should preserve heterogeneity where it is governance-relevant, avoid default majority averaging, and record contested dimensions as constraints on permissible use \citep{EspelandStevens2008,SelbstEtAl2019}.
Participation is treated as decision authority at defined gates rather than as post hoc feedback, and recourse includes the possibility of pausing, revising, or decommissioning systems when harms or misuse are documented \citep{Reisman2018,SelbstEtAl2019}.

\section{Case context and methods}
\label{sec:methods}

The empirical program is a three-year collaboration with community organizations and municipal partners in Montr\'eal, Canada.
Across both systems, participation is treated as co-production of evaluation criteria and governance artifacts, rather than as post hoc validation.
Imagery is used as the primary modality because it supports standardized stimuli and scalable inference, but the resulting artifacts are treated as partial measurement infrastructure rather than substitutes for lived experience \citep{Saha2019,Gabrys2016}.
The two systems serve different governance functions. \ProjA{} supports planning prioritization, comparative auditing, and post-intervention review under documented criteria; it is not authorized for individual-level decisions or enforcement. \ProjB{} supports early-stage ideation and structured deliberation over synthetic public-space imagery; it is not evidence of public support and it does not substitute for participatory design.

\subsection{Participation design}

Recruitment occurred through community partners to support inclusion of residents with distinct relationships to public space and to reduce barriers to participation.
Participants provided informed consent for the activities they joined and for any self-identification markers collected.
Participation was compensated, and engagements were designed with accessibility accommodations spanning in-person sessions and a web-based platform.

Data governance follows purpose limitation for civic uses.
Self-identification markers were used to study aggregation effects and to support disaggregated reporting, not to infer identity from imagery.
Across both systems, the intended downstream use is public-space planning and governance; data and outputs are not authorized for enforcement-adjacent repurposing.
Documentation is treated as a governance artifact that enables audit and contestation, including versioned criteria definitions, evaluation protocols, and records of disagreement \citep{nissenbaum-privacy,Gebru2021,Mitchell2019}.

\subsection{\ProjA{}: participatory assessment and scaling}

\ProjA{} integrates qualitative elicitation, structured evaluation, agreement analysis, and model-based scaling \citep{MushkaniKoseki2026StreetReview}.
Within PAU, it provides evidence about which criteria are benchmarkable, how subgroup differences surface under shared stimuli, and how far a co-produced measurement pipeline can be scaled before proxy limits become governance constraints.
Table~\ref{tab:study_design_overview} summarizes its modules and data artifacts.

\begin{table}[t]
\centering
\caption{\ProjA{} study design overview across integrated modules. In this paper, \ProjA{} functions as an empirical instantiation for PAU rather than as a standalone technical contribution; additional technical and methodological details are available in the Street Review article \citep{MushkaniKoseki2026StreetReview}.}

\label{tab:study_design_overview}
\small
\renewcommand{\arraystretch}{1.15}
\begin{tabularx}{\linewidth}{P{0.17\linewidth}P{0.23\linewidth}Y Y}
\toprule
\textbf{Module} & \textbf{Participants} & \textbf{Inputs and tasks} & \textbf{Primary outputs} \\
\midrule
Interviews and descriptor elicitation & 28 interviewees recruited via community organizations & Semi-directed interviews on inclusive streets; descriptor generation and thematic analysis & Resident-derived descriptors; consolidated criteria \\
\addlinespace
Focus groups and deliberative evaluation & 12 participants in 3 focus groups; additional ranking participants (N=17) & Individual and group scoring of street images; ranking experiment across 12 criteria; agreement analysis & Convergence and contestation measures; evidence on deliberation effects \\
\addlinespace
Scaling and citywide mapping & 12 evaluators for rating; model trained on co-produced labels & 120 rated images; label propagation across 15{,}000 local images; inference on 45{,}000 street images & Subgroup-aware predictions; interpretability analyses; citywide maps \\
\bottomrule
\end{tabularx}
\end{table}

\subsubsection{Descriptor elicitation and thematic consolidation}

\ProjA{} begins with semi-directed interviews that elicit resident descriptors of what makes streets inclusive or exclusionary.
Across 28 interviewees, participants produced more than 600 descriptors.
We consolidated descriptors through iterative thematic grouping into four criteria designed to remain interpretable across social positions and actionable in municipal routines: Accessibility, Aesthetics, Practicality, and Inclusivity.
Consolidation decisions were documented as a mapping from descriptors to criteria definitions to support later audit and revision.
Table~\ref{tab:criteria_definitions} summarizes the criteria and the types of cues participants discussed.

\begin{table}[t]
\centering
\caption{\ProjA{} criteria and indicative descriptors derived from participatory elicitation and thematic consolidation.}
\label{tab:criteria_definitions}
\small
\renewcommand{\arraystretch}{1.15}
\begin{tabular}{P{0.20\linewidth}P{0.74\linewidth}}
\toprule
\textbf{Criterion} & \textbf{Indicative descriptors and cues discussed by participants} \\
\midrule
Accessibility & Barrier-free routes, sidewalk continuity, manageable crossings, navigation ease, and conditions that enable mobility for people with varied physical abilities. \\
Aesthetics & Cleanliness, streetscape composition, greenery, and cues associated with maintenance. \\
Practicality & Signage and legibility, maintenance, availability of amenities, and cues associated with completing routine tasks. \\
Inclusivity & Safety and security cues, acceptance, cultural representation, and perceived belonging for different people. \\
\bottomrule
\end{tabular}
\end{table}

\subsubsection{Stimuli, evaluation protocol, and agreement}

For structured evaluation, \ProjA{} sampled 20 streets and represented them through 60 vantage points.
Participants first rated shared street images individually, then discussed the same stimuli in facilitated small groups, and finally recorded a collective group evaluation of those same streets.
The protocol therefore produces two linked but non-equivalent evidentiary objects: independent ratings and deliberative group judgments with associated reasons and residual disagreements.
Agreement was quantified using intra-class correlation coefficients for ratings and rank-based agreement measures for ranking tasks \citep{bartko1966,kooLi2016,puthSpearmanKendall2015}.

\subsubsection{Dataset construction}

The labeled scaling dataset contains 120 rated images.
Images are grouped into 60 street-level data-point groups corresponding to correlated viewpoints or nearby captures.
To prevent leakage and near-duplicate contamination, train, validation, and test splits are performed at the level of these groups rather than at the level of individual images \citep{RobertsEtAl2017}.
The scaling pipeline additionally uses a larger pool of local images for label propagation (15{,}000 images) and a broader set for inference (45{,}000 street images).
The resulting citywide map is an auditable representation with documented scope limits, not a claim to citywide ground truth.
All reported citywide outputs are therefore conditional on coverage and quality of the underlying imagery.

\subsubsection{Model family and outputs}

\ProjA{} trains a supervised multi-output regression model that predicts subgroup and aggregate ratings from street-view imagery.
The model predicts 28 scores: four criteria for each of six self-identified groups and four aggregate scores.
A semantic segmentation model extracts pixel-level features, and an attention-based multi-layer perceptron maps these features to predicted scores \citep{Vaswani2017}.
We report predictive performance using $R^2$ on validation and held-out test splits and treat the resulting maps as audit artifacts rather than as authoritative measurements.

\subsection{\ProjB{}: pluralistic preference data for generative governance}

\ProjB{} addresses the governance of synthetic public-space imagery used in planning and consultation \citep{MushkaniEtAl2025LIVS}.
Within PAU, it provides evidence about what preference tuning can and cannot justify when criteria are locally co-produced but optimization still requires reductions to a single policy.
It was developed through a two-year collaboration with community organizations.

\subsubsection{Criteria and prompts}

Building on previous work on criteria elicitation, participants in workshops and interviews elicited 634 concepts associated with equitable public-space design and consolidated them into six criteria: accessibility, safety, comfort, invitingness, inclusivity, and diversity. A prompting workshop produced 440 participant-authored prompts representing public-space scenarios.
The pipeline also includes prompts authored through structured augmentation to increase coverage of scenarios and criteria; prompt origin is retained as metadata for analysis.

\subsubsection{Image pipeline}

Using Stable Diffusion XL \citep{podellSDXL2023}, the pipeline generated 16{,}693 images.
Quality checks removed images with generation failures or obvious mismatches to the intended public-space setting, yielding 13{,}462 images for annotation.
Preference data were collected through an accessibility-oriented web-based platform.
Table~\ref{tab:livs_engagement} summarizes the co-production pipeline.

\begin{table}[t]
\centering
\small
\caption{Summary of \ProjB{} engagement activities and data outputs.}
\label{tab:livs_engagement}
\renewcommand{\arraystretch}{1.2}
\begin{tabularx}{\linewidth}{P{0.26\linewidth}P{0.30\linewidth}Y}
\toprule
\textbf{Engagement activity} & \textbf{Participants} & \textbf{Outputs} \\
\midrule
Criteria elicitation & Multiple workshops and interviews with community partners & 634 concepts consolidated into intermediate criteria and ranked to six final criteria: Accessibility, Safety, Comfort, Invitingness, Inclusivity, Diversity. \\
\addlinespace[0.5em]
Prompt authoring & 1 workshop, 24 participants & 440 human-authored prompts representing public-space scenarios and features. \\
\addlinespace[0.5em]
Annotations & 5 batches, 18 participants & 42{,}235 raw comparisons; 37{,}710 usable image-pair comparisons after removing quality-control items and incomplete submissions; 35{,}510 retained for DPO fine-tuning. \\
\addlinespace[0.5em]
Resident evaluation workshop & 1 workshop, 25 participants & 2{,}100 additional comparisons used for evaluation of fine-tuning and neutrality. \\
\bottomrule
\end{tabularx}
\end{table}

The preference task presented pairs of generated images and asked participants to select the image that better satisfied a criterion, with an explicit neutral option for indifference or indeterminacy.
The interface presented three criteria per image pair to reduce fatigue and reduce criterion stacking effects.
Filtering removed quality-control items, incomplete submissions, and image pairs tied to obvious generation failures or non-public-space mismatches. This yielded 37{,}710 usable image-pair comparisons, of which 35{,}510 were retained for DPO fine-tuning.
Because each usable comparison includes three criteria, the cleaned corpus corresponds to approximately 113{,}130 criterion-level responses. Neutral comparisons remain first-class outcomes for held-out evaluation and governance analysis even when they are not used as positive training signals.

\subsubsection{Preference tuning and evaluation}

A Direct Preference Optimization (DPO) fine-tune was performed on the retained subset \citep{rafailovDPO2024}.
Because DPO requires binary labels, training uses only non-neutral pairwise preferences, collapsed to a binary label by majority vote over the three criterion judgments shown for each pair; neutral selections are retained as governance-relevant outcomes in held-out evaluation.
Evaluation uses a held-out resident workshop dataset collected after fine-tuning (2{,}100 comparisons) and reports counts across outcomes and corresponding uncertainty estimates.
The held-out evaluation is separated in time from the training data collection and is used to assess whether preference tuning changes outcomes when the same neutral option is available.

\section{Findings}
\label{sec:findings}

This section synthesizes evidence from \ProjA{} and \ProjB{} into four propositions that connect empirical constraints to governance implications.
P1 uses descriptor frequencies, criterion correlations, and subgroup differences to test whether disagreement is structured; P2 uses the contrast between independent ratings and post-discussion group judgments; P3 uses grouped-split scaling performance and the 45{,}000-image representation as a bounded audit artifact; and P4 uses the held-out generative evaluation, neutrality-oriented diagnostics, and harm-focused review.
Table~\ref{tab:quantitative-summary} reports the quantitative results and study counts that are used in the propositions and in the governance requirements derived in \cref{sec:governance}.

\begin{table}[t]
\centering
\small
\caption{Quantitative summary of the evidence used in the propositions. Reported counts and metrics are drawn from the \ProjA{} and \ProjB{} pipelines described in \cref{sec:methods}.}
\label{tab:quantitative-summary}
\renewcommand{\arraystretch}{1.15}
\begin{tabularx}{\linewidth}{P{0.30\linewidth}P{0.22\linewidth}Y}
\toprule
\textbf{Artifact} & \textbf{Quantity or metric} & \textbf{Value} \\
\midrule
\ProjA{} stimuli & Streets; vantage points & 20 streets; 60 vantage points \\
\ProjA{} labeled set & Rated images; group-level split units & 120 images; 60 street-level data-point groups \\
\ProjA{} inference set & Images scored for mapping & 45{,}000 street images \\
\ProjA{} scaling performance & $R^2$ on validation; $R^2$ on held-out test & 0.91; 0.89 \\
\ProjA{} criterion correlations & Corr(Inclusivity, Accessibility); Corr(Inclusivity, Aesthetics); Corr(Practicality, Aesthetics) & 0.55; 0.54; $-0.05$ \\
\ProjB{} prompts and images & Participant-authored prompts; generated images; retained images & 440 prompts; 16{,}693 images; 13{,}462 images \\
\ProjB{} preference data & Raw comparisons; usable image-pair comparisons; DPO fine-tuning comparisons; criterion-level responses & 42{,}235; 37{,}710; 35{,}510; 113{,}130 \\
Post-tuning \ProjB{} evaluation & Comparisons; neutral share (95\% CI); DPO-preferred share (95\% CI) & 2{,}100; 52.4\% [50.2, 54.5]; 33.3\% [31.3, 35.3] \\
\bottomrule
\end{tabularx}
\end{table}

\subsection{P1: Pluralism is structured rather than noise}
\label{sec:p1}

Across \ProjA{}, pluralism appears as patterned variation rather than random disagreement. Participants produced more than 600 descriptors in elicitation, which consolidated into four criteria that remained distinguishable in subsequent evaluation. Quantitatively, correlations between criteria are moderate and uneven (Table~\ref{tab:quantitative-summary}): practicality and aesthetics are approximately uncorrelated ($-0.05$), while inclusivity correlates moderately with accessibility and aesthetics (0.55 and 0.54). In evaluation sessions, subgroup ratings on the same stimuli also differed across shared stimuli, underscoring that aggregation choices are consequential.

Pluralism in public-space governance has two implications for AI-mediated evidence.
First, a single composite score is not a neutral summary when criteria are not commensurable.
Second, aggregation choices can erase differences that matter for accountability when evaluations vary across social positions.

\ProjA{} addresses these implications by separating criteria and by producing subgroup-aware outputs rather than only a population-average score.
Even within the four \ProjA{} criteria, correlations are moderate and heterogeneous (Table~\ref{tab:quantitative-summary}).
In particular, practicality and aesthetics are approximately uncorrelated ($-0.05$), while inclusivity correlates moderately with accessibility and aesthetics (0.55 and 0.54).
These empirical patterns indicate that criteria capture distinct dimensions of evaluation and cannot be treated as interchangeable inputs without introducing discretionary trade-offs.
These relationships imply that treating criteria as interchangeable inputs to a single index would embed discretionary trade-offs that require public justification \citep{EspelandStevens2008,SelbstEtAl2019}.

On this basis, disagreement is treated as a structured property of the evaluation setup rather than as annotation error.
The governance requirement is to preserve and report heterogeneity where it is relevant to decision-making, and to restrict aggregation to cases where benchmarks are negotiated and documented in the \ValueRegister{}.

\subsection{P2: Deliberation changes the evidentiary object}
\label{sec:p2}

\ProjA{} collects both individual judgments and judgments produced after structured small-group discussion.
Empirically, these two stages produce different outputs even when applied to the same stimuli. In several cases, group discussions led participants to reinterpret criteria, attend to different visual cues, or revise earlier ratings. Group judgments were therefore not simple averages of individual scores, but reflected negotiated interpretations and recorded reasons.

Under the evidentiary-object framing, these outputs are not interchangeable because they are produced under different procedures and therefore support different forms of justification.
Deliberative evaluation can change how participants interpret criteria, how they weigh visible cues, and what reasons are recorded alongside scores \citep{Habermas1996,Fishkin2009}.

The protocol therefore yields two empirically distinct artifacts: independent ratings and deliberative judgments that include reasons and residual disagreement.
Within PAU, deliberation is not treated as a method for forcing consensus.
Instead, deliberative sessions are designed to produce a record of reasons and residual disagreement that can be used to determine whether a criterion is treated as benchmarkable, treated as contested, or treated as out of scope for the modality.

Residual disagreement after deliberation is recorded in the \ValueRegister{} and constrains permissible uses of models and maps, rather than being removed through forced aggregation \citep{Mouffe2000,EspelandStevens2008}.

\subsection{P3: Co-produced scaling is feasible but bounded}
\label{sec:p3}

\ProjA{} demonstrates that co-produced judgments can support predictive scaling and citywide representation. Using group-level splits that prevent near-duplicate leakage, the scaling model attains $R^2=0.91$ on validation and $R^2=0.89$ on a held-out test set. Applied to 45{,}000 street images, it yields a citywide representation that can support auditing and prioritization under documented scope limits. Because the pipeline operates on street-view imagery, that representation is necessarily tied to visible streetscape features rather than to broader experiential conditions \citep{Alpherts2024Perceptive}.

The same evidence indicates bounded feasibility.
The input modality constrains what the model can represent.
Cultural and symbolic markers of belonging, and experiential factors such as harassment or policing, are not consistently visible in street imagery.
Data quality and coverage also vary across neighborhoods, and model outputs inherit these gaps.
Subgroup modeling relies on coarse self-declared categories that are used to reveal majority averaging but can reify groups and obscure within-group heterogeneity.
These observed limitations arise directly from the data modality, coverage, and labeling scheme used in the pipeline.
These limitations imply that maps should be treated as evidence subject to audit and contestation, not as authoritative measures \citep{Star1999,Gebru2021,Mitchell2019}.

\subsection{P4: Neutrality in preference-based alignment is a governance signal}
\label{sec:p4}

\ProjB{} shows that preference-based evaluation of generated public-space imagery produces indeterminacy when the evaluation task does not support a determinate ordering. In the resident evaluation workshop with 2{,}100 new comparisons, participants favored DPO-tuned outputs in 700 instances, favored the baseline in 300 instances, and selected neutral in 1{,}100 instances (Table~\ref{tab:bootstrap-cis}). The neutral share is 52.4\% (bootstrap 95\% CI [50.2, 54.5]). Conditioning on non-neutral outcomes only, DPO-tuned outputs are preferred in 70.0\% of cases (bootstrap 95\% CI [67.1, 72.8]).

\begin{table}[t]
\centering
\small
\caption{Outcome proportions in the \ProjB{} evaluation workshop (2{,}100 comparisons) with bootstrap 95\% confidence intervals computed by multinomial resampling.}
\label{tab:bootstrap-cis}
\renewcommand{\arraystretch}{1.15}
\begin{tabular}{lccc}
\toprule
\textbf{Outcome} & \textbf{Count} & \textbf{Proportion} & \textbf{Bootstrap 95\% CI} \\
\midrule
DPO-tuned preferred & 700 & 0.333 & [0.313, 0.353] \\
Baseline preferred & 300 & 0.143 & [0.128, 0.158] \\
Neutral & 1{,}100 & 0.524 & [0.502, 0.545] \\
\bottomrule
\end{tabular}
\end{table}

Neutral responses are not randomly distributed: they are more likely in cases where images are visually similar, prompts are underspecified, or criteria require context not present in the image.
Under PAU, neutrality is treated as evidence that the comparison task does not support a stable ordering under the criterion, either because the prompt or imagery under-specifies relevant context or because values conflict in ways not captured by pairwise ordering.
Neutrality therefore constrains what preference tuning can justify.

The governance requirement is to treat neutrality rates and harm flags as triggers for review, scope narrowing, or alternative engagement methods, rather than as labels to be smoothed through aggregation \citep{sorensen2024roadmap,Conitzer2024}.

\subsection{Limits on use}
\label{sec:failure-modes}

Across the two systems, failures and indeterminacy correspond to governance-relevant constraints.
For \ProjA{}, the modality emphasizes visible design cues and limits representation of experiential harms and symbolic recognition.
For \ProjB{}, preference tuning changes outcomes when participants make a non-neutral choice, but the high neutral rate indicates that many comparisons do not support determinate ordering.
Harm-focused review further targets failure modes that are not captured by scalar criterion preferences.
For example, prompts involving safety can elicit images that insert police or security presence as a proxy for safety, which changes the meaning of safety for groups that experience differential enforcement \citep{Brayne2017,Eubanks2018}.
In PAU, these patterns are treated as evidence that constrains permissible use, including explicit scope limits and pause authority in the governance architecture.

\section{Governance and implementation architecture}
\label{sec:governance}

The propositions imply that legitimacy requires procedures that preserve heterogeneity, document disagreement and indeterminacy, and provide standing to contest how AI-mediated evidence is produced and used.
This section translates the propositions into a municipal governance architecture that can be attached to procurement and recurring stages of development and maintenance.

\subsection{Translating findings}

Table~\ref{tab:requirements-p1-p4} links empirically grounded conditions to governance requirements and primary failure modes.

\begin{table}[t]
\centering
\small
\caption{Governance requirements derived from propositions P1 to P4. The table links empirically grounded constraints to procedural implications for municipal governance.}
\label{tab:requirements-p1-p4}
\renewcommand{\arraystretch}{1.15}
\begin{tabularx}{\linewidth}{P{0.10\linewidth}P{0.24\linewidth}Y P{0.24\linewidth}}
\toprule
\textbf{Prop.} & \textbf{Empirical condition} & \textbf{Governance requirement} & \textbf{Primary failure mode if ignored} \\
\midrule
P1 & Criteria are not commensurable and aggregation can erase heterogeneity & Preserve subgroup-aware evidence; avoid default majority averaging; maintain a \ValueRegister{} that separates benchmarks from contested dimensions \citep{crenshaw1989,barocas2016disparate}. & Universal metric treated as legitimate; harms erased through commensuration; governance capture by dominant norms \citep{EspelandStevens2008,Strathern2000}. \\
\midrule
P2 & Procedure changes the evidentiary object, including what reasons and disagreements are recorded & Institutionalize \DelibCells{}; maintain disagreement logs; treat residual disagreement as evidence that constrains permissible uses \citep{dewey1927public,Habermas1996,Healey1997}. & Technocratic closure through forced consensus or administrative paralysis without negotiation mechanisms. \\
\midrule
P3 & Scaling is feasible but bounded by modality limits and uneven data quality & Constrain permissible uses; require auditability and documentation; incorporate data quality checks and drift monitoring; ensure recourse for systematic misrepresentation \citep{Sculley2015,Tabassi2023,Ibrahim2020}. & Overtrust in maps and scores; mission creep; uneven visibility produces uneven governance capacity \citep{Star1999,Kitchin2016}. \\
\midrule
P4 & Indeterminacy persists in preference evaluation and harm-focused review & Treat neutrality and harm flags as triggers for review or scope narrowing; avoid optimization toward a single normative target & False precision; procedural illegitimacy; alignment claims used to justify contested decisions without deliberation \citep{Pasquale2015,Merry2016}. \\
\bottomrule
\end{tabularx}
\end{table}

\subsection{Lifecycle governance}

PAU uses an augmented participatory AI lifecycle with five phases: co-framing, co-design, co-implementation, co-deployment, and co-maintenance.
The phases correspond to problem definition and authorization, criteria and protocol design, data and model development, pilot and operational use, and revision or sunset under monitoring.
Lifecycle governance treats updates and maintenance as moments where values are re-encoded, so it defines decision gates, required artifacts, and pause authority \citep{Jasanoff2004,Sculley2015}.
Table~\ref{tab:workflow-checkpoints} maps checkpoints to phases.

\begin{table}[t]
\centering
\small
\caption{Municipal workflow checkpoints and pause authority mapped to an augmented participatory lifecycle for commissioned public-space AI systems.}
\label{tab:workflow-checkpoints}
\renewcommand{\arraystretch}{1.15}
\begin{tabularx}{\linewidth}{P{0.18\linewidth}P{0.20\linewidth}Y P{0.28\linewidth}}
\toprule
\textbf{Phase} & \textbf{Decision gate} & \textbf{Required artifacts (minimum)} & \textbf{Pause trigger} \\
\midrule
Co-framing & Authorization to proceed to design & Project charter; \ValueRegister{} v0; risk register; documented decision boundaries & Contested values collapsed into a single metric; disallowed uses not constrained \\
\midrule
Co-design & Authorization to implement and procure & Data governance plan; evaluation protocol including indeterminacy; communication plan; oversight and recourse plan & Missing stewardship provisions; no disaggregation plan; no recourse pathway \\
\midrule
Co-implementation & Authorization to pilot & Model card; data documentation; baseline evaluation and error analysis; traceability mapping & Error rates unacceptable for identified populations or geographies; non-auditable pipeline \\
\midrule
Co-deployment & Authorization to continue or scale & Pilot memo; public notice; logging infrastructure; assessment report & Use beyond approved scope; complaint clusters indicating systematic harm; failure to meet assessment conditions \\
\midrule
Co-maintenance & Authorization to renew, revise, or sunset & Versioned documentation; drift report; re-commissioning report; updated \ValueRegister{} & Detected drift or coverage changes; scope expansion; sustained indeterminacy; governance non-compliance \\
\bottomrule
\end{tabularx}
\end{table}

\subsection{Procurement requirements}
Municipal procurement is a recurrent point at which standards, audit access, and scope limits can be enforced. We therefore specify commissioning clauses that operationalize the \RightToAI{} and lifecycle governance; the full template appears in Appendix~\ref{app:procurement}.

\subsection{Oversight topology and recourse}
\begin{figure}[t]
\centering
\begin{adjustbox}{max width=\linewidth}
\begin{tikzpicture}[
font=\footnotesize,
>={Latex},
box/.style={
draw, rounded corners=3pt, align=center,
inner xsep=7pt, inner ysep=6pt,
text width=0.195\linewidth,
minimum height=27mm
},
sbox/.style={
box,
text width=0.185\linewidth,
minimum height=25mm
},
arrow/.style={->, line width=0.85pt},
dashedarrow/.style={->, line width=0.85pt, dashed}
]

\node[box] (proj) {%
{\bfseries Project team}\\
{\scriptsize Planning, design, operations\\
Implements, logs, responds}%
};

\node[box, right=12mm of proj] (ai) {%
{\bfseries Department AI and ethics}\\
{\scriptsize Compliance, documentation\\
Audit readiness}%
};

\node[box, right=12mm of ai] (cell) {%
{\bfseries \DelibCell{}}\\
{\scriptsize Reviews contested dimensions\\
Updates \ValueRegister{}\\
Recommends pause or revision}%
};

\node[box, right=12mm of cell] (council) {%
{\bfseries Council committee}\\
{\scriptsize Authorizes high-impact use\\
Approves scope changes}%
};

\node[sbox, below=12mm of proj] (public) {%
{\bfseries Public}\\
{\scriptsize Notice and participation\\
Complaints and recourse}%
};

\node[sbox, below=12mm of cell] (omb) {%
{\bfseries Independent office}\\
{\scriptsize Appeals\\
Independent audits}%
};

\draw[arrow] (proj.east) to[out=8, in=172] (ai.west);
\draw[arrow] (ai.east)   to[out=8, in=172] (cell.west);
\draw[arrow] (cell.east) to[out=8, in=172] (council.west);

\draw[dashedarrow] (public.north) to[out=90, in=260] (proj.south);

\draw[dashedarrow]
([yshift=3mm]public.east)
to[out=0, in=210] ([xshift=-2mm,yshift=-4mm]cell.west);

\draw[dashedarrow]
([yshift=-3mm]public.east)
to[out=0, in=180] (omb.west);

\draw[arrow] (omb.north) to[out=90, in=270] (cell.south);

\draw[arrow]
(omb.east)
to[out=0, in=270] ([xshift=3mm]council.south);

\end{tikzpicture}
\end{adjustbox}
\caption{Oversight topology for municipal public-space AI governance. \DelibCells{} provide recurring participatory oversight connected to departmental compliance functions and democratic authorization, with an independent recourse pathway.}
\Description{Flow diagram showing an oversight structure linking a project team, departmental AI and ethics function, a deliberative cell, and a council committee, with complaint and appeals pathways from the public.}
\label{fig:oversight-topology}
\end{figure}
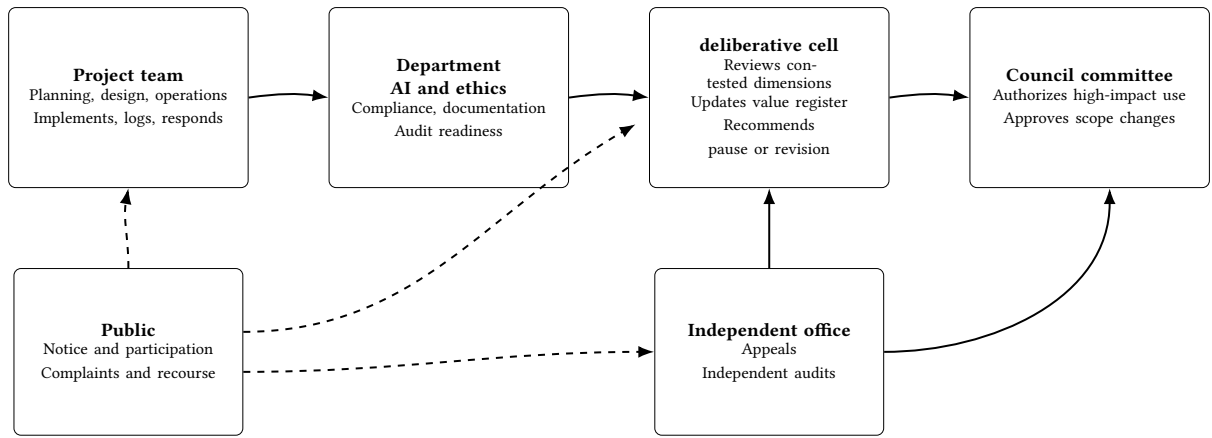

PAU assigns responsibility across delivery, compliance, participatory review, and democratic authorization.
Figure~\ref{fig:oversight-topology} presents an oversight topology that connects project teams, departmental AI and ethics functions, \DelibCells{}, and council committees, with a public complaint pathway and an independent appeals option.
\DelibCells{} have standing to update the \ValueRegister{} and to recommend revision or pause when disagreement, drift, misuse, or harm flags are detected \citep{Fung2004,Reisman2018}.
Operationally, the mapping in Figure~\ref{fig:oversight-topology} is actor-artifact-action: project teams maintain model cards, logs, and incident reports; departmental AI and ethics functions verify documentation and procurement compliance; \DelibCells{} update the \ValueRegister{} and disagreement log and can recommend pause or rollback; council committees authorize scope changes; and residents initiate review through complaints and appeals.

\subsection{Aggregation and disallowed uses}
\label{sec:scope-limits}

The architecture imposes scope limits derived from pluralism and modality bounds.
For representational systems, aggregation is treated as a discretionary choice that requires justification.
When agreement is high and a benchmark is negotiated, aggregation can support budgeting and prioritization.
When a dimension is contested or weakly supported by the modality, outputs remain disaggregated and accompanied by documented disagreement and uncertainty.

Table~\ref{tab:allowed-disallowed-uses} provides testable scope limits for representational and generative outputs.
The scope limits are framed as conditions for use and conditions for re-authorization, rather than as aspirational principles.

\begin{table}[t]
\centering
\small
\caption{Scope limits and disallowed uses for public-space representational and generative systems under the procedural \RightToAI{}.}
\label{tab:allowed-disallowed-uses}
\renewcommand{\arraystretch}{1.15}
\begin{tabularx}{\linewidth}{P{0.26\linewidth}Y Y}
\toprule
\textbf{Output type} & \textbf{Authorized uses (within scope)} & \textbf{Disallowed uses or re-authorization required} \\
\midrule
Representational scores and maps & Planning, design prioritization, and auditing of spatial inequities under documented criteria; disaggregated reporting when criteria are contested. & Individual-level decisions; enforcement, eligibility, or sanctioning; repurposing as surveillance justification; use without documentation of criteria and aggregation choices. \\
\midrule
Generated public-space imagery & Early-stage ideation, communicating alternative design features, and supporting structured deliberation when labeled as synthetic and accompanied by neutrality and harm screening. & Evidence of public support; substitution for participatory design; use in high-stakes safety justification without independent evidence; generation or deployment of depictions that trigger harm flags without documented review and mitigation. \\
\bottomrule
\end{tabularx}
\end{table}

\noindent\textbf{Example.} To illustrate how scope limits, authorization gates, and pause and rollback mechanisms operate in practice, we present an end-to-end municipal vignette of a corridor redesign in Appendix~\ref{app:vignette}.

\section{Discussion}
\label{sec:discussion}

PAU frames fairness and accountability in public-space AI systems as problems of institutional legitimacy under pluralism.
For AI accountability and governance research, the results show that neutrality rates and non-commensurable criteria correlations are properties of the evidentiary object and therefore inputs to governance.
For HCI and participatory design, the systems illustrate how participation can be attached to criteria formation, evaluation protocols, and documentation artifacts that structure future contestation \citep{costanzaChock2020}.
For urban planning, the work treats AI-mediated representations and generative imagery as infrastructure through which streets are governed, so procurement and maintenance become sites where values are encoded and must remain revisable \citep{Star1999,Mattern2017CityNotComputer}.

\subsection{Addressing the gap identified in related work}
The gap in Section~\ref{sec:related} concerned how municipalities might treat disagreement and indeterminacy as decision-relevant evidence once AI outputs begin to function as civic evidence. PAU can be read as one way of addressing this gap by linking empirical constraints to administrative artifacts and levers: criteria and subgroup differences inform the \ValueRegister{}; deliberative sessions produce disagreement logs and benchmark decisions; scaling results are translated into scope limits on maps and scores; and neutrality in generative evaluation serves as a trigger for review rather than something to be optimized away. Taken together, this suggests a governance architecture that helps clarify what public-space AI might reasonably justify, not only what it can predict or generate.

\subsection{Future directions}
Three directions follow from the cases. First, PAU could be re-commissioned across cities rather than transferred as a fixed template. This would help examine which criteria travel, which remain local, and how often municipal value registers require revision as demographics and policies change. Second, public-space governance would likely benefit from stronger evidence pathways for harms that are not well expressed in images, including harassment, symbolic exclusion, and temporal use patterns. In PAU terms, this points to combining image-based artifacts with other modalities rather than relying on vision systems alone. Third, future work could further examine institutional robustness and governance validity: for example, whether deliberative cells remain representative over time, whether pause and rollback powers are actually used, and which procurement clauses municipalities are able to sustain as systems are updated, re-scoped, or contested in practice.

\subsection{Limitations}
\label{sec:limitations}

Internal validity threats concern whether measured outcomes reflect the intended constructs.
For \ProjA{}, correlated imagery can produce leakage if train and test splits are not performed at the street-level group, and the pipeline therefore enforces group-level splitting.
Deliberative evaluation can be affected by group dynamics and by fatigue; PAU treats deliberation as procedure-dependent evidence rather than as a source of ground truth.
For \ProjB{}, neutral selection can reflect ambiguity or fatigue in addition to indeterminacy; the platform logs response times and retains prompt-origin and similarity metadata to support neutrality interpretation.

External validity threats concern how results generalize beyond this setting.
Both systems are grounded in Montr\'eal and one set of community partners, so the distributions of descriptors and the salience of criteria are not assumed to transfer without re-commissioning.
Both systems rely on imagery, which limits representation of experiential and symbolic dimensions, and which inherits spatial coverage differences in the underlying image sources.

Governance validity threats concern whether institutions can implement the procedures in ways that preserve pluralism rather than erode it.
The architecture presumes municipal willingness to allocate authority and budget to participation and oversight, and it can be undermined by capture or by procedural compliance without substantive responsiveness.
The templates in Appendix~\ref{app:templates} are designed to make decision boundaries and revision authority explicit, but they do not eliminate the political work required to enforce scope limits.

\section{Conclusion}
\label{sec:conclusion}

Public-space AI systems do not simply describe urban conditions; they participate in how those conditions are defined, compared, and acted upon. As scores, maps, and synthetic images enter budgeting, prioritization, and public justification, the assumptions embedded in criteria, aggregation, and evaluation protocols become questions of governance rather than technical detail.

This paper proposed Pluralistic-Alignment Urbanism (PAU) as a way of governing these systems under conditions of persistent disagreement. Across two participatory case studies, we showed that disagreement is structured rather than random, that deliberation changes what counts as evidence, that co-produced judgments can be scaled but only within clear modality limits, and that neutrality in preference-based evaluation constrains what generative systems can justify.

Rather than treating these features as obstacles to be optimized away, PAU treats them as inputs to governance. The framework translates them into procedures and artifacts that preserve heterogeneity, document disagreement, and define when AI-mediated representations can and cannot support municipal decisions. This includes disaggregated reporting, versioned value registers, standing deliberative bodies, and explicit pause and rollback authority.

The central claim is not that municipal AI can resolve disagreement about public space, but that it should be governed in ways that keep such disagreement visible, contestable, and consequential. Under this view, the legitimacy of public-space AI systems depends less on predictive performance alone and more on whether the institutions surrounding them can sustain ongoing revision, contestation, and scope control as urban conditions and values change.

\section*{Ethical Review and Human Subjects}
This study involving human participants was reviewed and approved by the authors’ institutional Research Ethics Board. All participants provided informed consent, participation was voluntary and compensated, and data were used only for the purposes described in this research.

\section*{Generative AI Usage Statement}
Generative AI tools were used for grammar and stylistic editing. All ideas, analyses, and written content are the authors’ own; no generative AI was used for data generation, analysis, or interpretation.

\bibliographystyle{ACM-Reference-Format}
\bibliography{0_01_references}
\clearpage
\appendix

\section{Governance templates and documentation artifacts}
\label{app:templates}

This appendix provides implementable templates for the governance artifacts referenced in the paper.
All templates are proposed to be (i) versionable, (ii) auditable, and (iii) usable as procurement attachments.

\subsection{Public Space AI Charter template}

\noindent\textbf{Purpose.} A one-page artifact that authorizes (or declines to authorize) a system for a specified municipal use.

\noindent\textbf{Template fields (minimum).}
\begin{itemize}[leftmargin=1.5em]
  \item \textbf{System name and owner:} department, program, accountable official.
  \item \textbf{Intended use:} decisions the system may inform; time horizon; deployment context.
  \item \textbf{Decision boundary statement:} explicit \emph{disallowed uses} (e.g., enforcement-adjacent repurposing; individual-level decisions).
  \item \textbf{Evidence modality and scope limits:} what the system can/cannot represent (e.g., imagery limits on experiential harms).
  \item \textbf{Public notice plan:} how and where notice is published; how to obtain documentation.
  \item \textbf{Participation and authority:} which engagements occur; who has standing; what can be revised.
  \item \textbf{Recourse path:} complaint intake channels; response SLA; escalation/appeals option.
  \item \textbf{Versioning:} links/pointers to the current \ValueRegister{} version and model/data documentation.
\end{itemize}

\subsection{Value Register template}

\noindent\textbf{Purpose.} A versioned register that distinguishes benchmarkable dimensions from contested ones and encodes scope limits.

\noindent\textbf{Template fields (minimum).}
\begin{enumerate}[leftmargin=1.8em]
  \item \textbf{Metadata:} \ValueRegister{} version ID; adoption date; adopting body (e.g., \DelibCell{}); next review date.
  \item \textbf{System scope:} covered neighborhoods/areas; modality; known exclusions.
  \item \textbf{Criteria definitions:} each criterion with (i) definition, (ii) intended municipal interpretation, (iii) non-goals.
  \item \textbf{Status per criterion:} \emph{Benchmarkable}, \emph{Contested}, or \emph{Out-of-scope for modality}.
  \item \textbf{Disaggregation rules:} which subgroup breakdowns are required; when aggregation is prohibited.
  \item \textbf{Aggregation rules (if any):} allowed composite metrics; weighting rationale; validity conditions.
  \item \textbf{Indeterminacy handling:} how neutrality/disagreement is reported; thresholds that trigger review.
  \item \textbf{Authorized uses / disallowed uses:} testable conditions for use; re-authorization triggers.
  \item \textbf{Lifecycle checkpoints:} required artifacts at each gate; pause/rollback authority and triggers.
\end{enumerate}

\subsection{Disagreement log template}

\noindent\textbf{Purpose.} A durable record of unresolved value conflict and deliberative reasons that constrain permissible use.

\noindent\textbf{Template fields (minimum).}
\begin{itemize}[leftmargin=1.5em]
  \item Meeting/session ID; date; facilitation method; participant composition (high level, privacy-preserving).
  \item Issue statement (what is contested).
  \item Positions summarized (at least two distinct positions if present).
  \item Evidence cited (artifacts, examples, maps, prompts, outputs).
  \item Residual disagreement (what did not resolve) and \emph{why}.
  \item Decision (if any): revise criteria / restrict use / collect new evidence / pause.
  \item Action items: owner, deadline, and the artifact(s) to be updated (e.g., \ValueRegister{} vX).
\end{itemize}

\subsection{Pause and rollback memo template}

\noindent\textbf{Purpose.} A standardized memo that operationalizes pause authority and makes rollback traceable.

\noindent\textbf{Template fields (minimum).}
\begin{itemize}[leftmargin=1.5em]
  \item Trigger category: drift, misuse, complaint cluster, harm flags, coverage change, governance non-compliance.
  \item Evidence summary: logs, audits, complaints, evaluation results (include version identifiers).
  \item Immediate mitigation: suspend outputs, revert to prior version, restrict scope, increase notice.
  \item Rollback plan: which version is restored; what downstream artifacts are corrected.
  \item Re-commissioning requirements: what must be redone before reactivation (data, protocol, deliberation).
  \item Public communication: what is disclosed, when, and where; how residents can contest or contribute.
\end{itemize}

\subsection{Neutrality validity checks and neutrality predictors}

Because a neutral option can reflect indifference, ambiguity, fatigue, or value conflict, \ProjB{} included design features intended to support interpretation.
The evaluation protocol includes an explicit statement that neutrality can mean indifference, insufficient visual evidence, or inability to decide under the criterion.
The platform logs response time per comparison to support fatigue checks.
The prompting pipeline records whether a prompt is participant-authored or pipeline-authored to support ambiguity analysis, because prompts constructed by automated augmentation can omit context that affects criteria such as Inclusivity and Diversity.
Pairs are associated with image similarity proxies derived from embedding similarity, which is used to assess whether neutrality is concentrated on near-ties.

For neutrality diagnostics, we specify a mixed-effects logistic regression in which the dependent variable is whether the response is neutral and predictors include criterion, prompt origin, similarity proxy, and response-time quantiles as a fatigue proxy. Random intercepts for annotators and prompts account for repeated measures and prompt-specific difficulty. The purpose is to distinguish neutrality plausibly attributable to similarity, ambiguity, or fatigue from neutrality that remains and therefore constrains what preference data can justify.

This appendix specifies the mixed-effects model used to structure analysis of neutral selections in \ProjB{}.

\paragraph{Outcome.}
Let $i$ index a criterion-level response (a single criterion judgment for a displayed image pair).
Define $y_i = 1$ if the annotator selected \emph{Neutral} for response $i$, and $y_i = 0$ otherwise.

\paragraph{Covariates (fixed effects).}
\begin{itemize}[leftmargin=1.5em]
  \item \textbf{Criterion} (categorical): Accessibility, Safety, Comfort, Invitingness, Inclusivity, Diversity.
  \item \textbf{Prompt origin} (binary): participant-authored vs pipeline-authored/augmented.
  \item \textbf{Similarity proxy} (continuous): embedding similarity (higher indicates closer pair).
  \item \textbf{Response-time quantile} (categorical): e.g., quintiles of response time per annotator as a fatigue proxy.
\end{itemize}

\paragraph{Model.}
\begin{equation}
\label{eq:neutrality-glmm}
y_i \sim \mathrm{Bernoulli}(\pi_i), \qquad
\mathrm{logit}(\pi_i) =
\beta_0
+ \beta_{\mathrm{crit}[i]}
+ \beta_{\mathrm{orig}}\,\mathrm{Orig}_i
+ \beta_{\mathrm{sim}}\,\mathrm{Sim}_i
+ \beta_{\mathrm{rt}[i]}
+ u_{\mathrm{annot}[i]}
+ v_{\mathrm{prompt}[i]}.
\end{equation}

\noindent Random intercepts:
\[
u_{\mathrm{annot}[i]} \sim \mathcal{N}(0,\sigma^2_{\mathrm{annot}}),
\qquad
v_{\mathrm{prompt}[i]} \sim \mathcal{N}(0,\sigma^2_{\mathrm{prompt}}).
\]

\paragraph{Interpretation under PAU.}
High neutrality that persists after accounting for similarity and response-time is treated as a governance constraint on what the preference task can justify (scope narrowing, deliberation, or alternative evidence modalities), rather than as an error term to be optimized away.

\subsection{Harm-focused generative evaluation protocol}

The generative system is evaluated not only for criterion satisfaction but also for representational harm risks that can arise when imagery enters civic settings.
The harm-focused protocol targets stereotype and tokenism failure modes around disability, race, and gender expression, and it targets policing and surveillance cues that can change perceived welcome and safety \citep{Noble2018,BenjaminRhua2019,Crawford2021}.
The protocol combines prompt-level checks, image-level coding, and governance interpretation.
Prompt-level checks identify prompts that specify demographic attributes or safety settings.
Image-level coding flags demographic skew, tokenistic single-person inclusion, stereotyped roles, and enforcement-adjacent cues.
Governance interpretation treats these flags as constraints on permissible use and as inputs to deliberation, rather than as errors to be minimized through optimization.

Furthermore, this appendix summarizes the harm-focused coding scheme used to screen \ProjB{} generated public-space images for representational failure modes that are not well-captured by scalar criterion preferences.

\begin{table}[t]
\centering
\small
\caption{Harm-focused coding scheme (summary). Codes are intended as governance flags and review triggers, not as optimization targets.}
\label{tab:harm-codes}
\renewcommand{\arraystretch}{1.15}
\begin{tabularx}{\linewidth}{P{0.18\linewidth}Y P{0.20\linewidth}}
\toprule
\textbf{Code} & \textbf{Definition (what is flagged)} & \textbf{Governance action} \\
\midrule
Enforcement proxy shift & Safety depicted primarily through police/security presence, weapons, or coercive enforcement cues that can change the meaning of ``safety'' across groups. & Review; restrict use in consultation; potential prompt class pause. \\
\midrule
Surveillance escalation & Prominent CCTV, monitoring towers, facial recognition-like framing, or pervasive monitoring aesthetics without explicit authorization context. & Review; require explicit labeling and deliberation; potential scope restriction. \\
\midrule
Tokenistic inclusion & Single visible ``representative'' person used as a diversity cue without context; inclusion as a prop rather than a plausible scene. & Review; do not use as exemplars; revise prompt set. \\
\midrule
Stereotyped roles & Depictions that place demographic groups into stereotyped social roles (e.g., disability only in dependency frames). & Review; remove from engagement set; prompt revision. \\
\midrule
Demographic skew & Systematic absence or overrepresentation of apparent demographics across a prompt class inconsistent with civic intent. & Aggregate audit; re-balance prompt set; re-commission if persistent. \\
\midrule
Context failure & Scene is not plausibly a public-space setting or introduces unrelated hazards that distort interpretation. & Filter; do not present in civic materials. \\
\bottomrule
\end{tabularx}
\end{table}

\paragraph{Coding notes.}
Codes are based on \emph{apparent visual cues} and should be applied conservatively.
Ambiguous cases are routed to deliberative review rather than forced into a definitive label.

\subsection{Procurement and commissioning clauses}
\label{app:procurement}

\noindent\textbf{Purpose.} Minimum contractual provisions that operationalize the \RightToAI{} and lifecycle governance in municipal commissioning.

\begin{table}[t]
\centering
\small
\caption{Procurement and commissioning clauses that operationalize the \RightToAI{} and lifecycle governance for public-space AI systems and AI-mediated street projects.}
\label{tab:procurement-clauses}
\renewcommand{\arraystretch}{1.15}
\begin{tabularx}{\linewidth}{P{0.26\linewidth}Y}
\toprule
\textbf{Clause category} & \textbf{Minimum contract requirements} \\
\midrule
Disclosure and boundaries & Intended uses and disallowed uses; decision boundary statement; public notice; prohibition on enforcement-adjacent repurposing without explicit authorization and safeguards. \\
\midrule
Auditability and access & Model card and data documentation; evaluation results and update logs; independent audit rights; logging of outputs and downstream uses; access to versioned documentation. \\
\midrule
Data stewardship & Consent and retention terms; restrictions on secondary use; governance of participatory data; privacy safeguards where applicable; compensation and accessibility supports for participation. \\
\midrule
Plural representation & Evaluation plan specifying where disaggregation is required; explicit handling of disagreement and indeterminacy; documentation of contested dimensions and associated safeguards. \\
\midrule
Re-commissioning and sunset & Update schedule and costs; triggers for pause and rollback; re-authorization requirements for scope changes; decommissioning and data retirement plan. \\
\bottomrule
\end{tabularx}
\end{table}

\section{Example municipal vignette: corridor redesign with pause and rollback}
\label{app:vignette}

This vignette illustrates how PAU attaches procedures to an end-to-end municipal workflow.
The setting is a corridor redesign project in which a planning department intends to use a streetscape scoring system to prioritize segments for accessibility upgrades and intends to use synthetic imagery to support public engagement about design alternatives.
The vignette is representative in its sequence of actions and artifacts, not in its location-specific details.

\paragraph{Co-framing and authorization.}
A project team proposes a corridor redesign and requests authorization to use a streetscape scoring system to support segment prioritization.
Under PAU, authorization requires a project charter that specifies the decisions the system will inform, the time horizon of use, and a decision boundary statement clarifying what the system will not be used for.
The charter attaches a \ValueRegister{} v0 listing proposed criteria, proposed aggregation rules, and the anticipated downstream decisions.
At this gate, a \DelibCell{} reviews whether any proposed criteria are contested or weakly supported by imagery and requires disaggregated reporting for those dimensions.
The \DelibCell{} also records disallowed uses in the \ValueRegister{}, including any enforcement-adjacent repurposing.

\paragraph{Co-design of criteria, evaluation protocol, and communication.}
The project team convenes participatory sessions to elicit criteria and to refine definitions.
For \ProjA{}-style representational outputs, the resulting criteria definitions and the mapping from descriptors to criteria are recorded as versioned artifacts.
The evaluation protocol specifies whether judgments are elicited individually, deliberatively, or both, and it specifies how disagreement will be recorded.
A communication plan specifies how maps will be presented in public materials, including an explanation of what the model can and cannot represent.
For synthetic imagery, the co-design phase specifies the prompts to be used for engagement, the criteria under which images will be assessed, and the conditions under which neutrality will trigger additional deliberation rather than selection of a single image as determinately preferred.

\paragraph{Co-implementation, baseline evaluation, and procurement.}
Before deployment, the vendor or internal team produces a model card, data documentation, and baseline evaluation results.
The project team assesses the intended use against the \ValueRegister{}.
If the corridor has known imagery coverage gaps, the team documents the expected spatial blind spots and specifies how those gaps affect the use of the map for prioritization.
For synthetic imagery, the project team records the base model, the fine-tuning procedure, and a harm-focused screening protocol.
Procurement clauses require audit access, logging, and update notifications, and they specify that any material changes to criteria or model behavior require re-commissioning.

\paragraph{Co-deployment and decision use.}
In the pilot phase, the project team uses disaggregated \ProjA{} outputs to identify segments with consistently low accessibility ratings and to flag segments where predicted evaluations differ across groups.
The team uses these outputs as one input to prioritization, alongside engineering constraints and budget.
When presenting results to the public, the team includes the \ValueRegister{} version, the decision boundary statement, and a disagreement log that documents contested dimensions.
For engagement materials, the team uses synthetic imagery only after labeling it as synthetic, attaching the prompts used, and reporting neutrality outcomes from the evaluation protocol.
If neutrality rates are high for a prompt class, the team treats this as a signal that imagery is not supporting determinate comparison for that criterion and shifts to alternative evidence, such as site visits or facilitated discussion without images.

\paragraph{Pause and rollback triggers in co-maintenance.}
During operation, the logging infrastructure records which maps and images were used in which decision contexts.
If public complaints cluster around a neighborhood not well covered by the imagery, or if residents report that a map systematically misrepresents their experiences, the \DelibCell{} can recommend a pause.
For representational systems, pause authority is triggered by documented drift, coverage changes, or persistent contested dimensions that are being used as benchmarks without authorization.
For synthetic imagery, pause authority is triggered when harm-focused screening flags outputs that shift safety meaning through enforcement cues, or when synthetic images are treated as evidence of support.
A pause results in a documented rollback to prior versions or a suspension of use until re-commissioning completes.
The recourse record, audit log, and \ValueRegister{} are updated to capture the incident and the corrective actions.

\end{document}